\documentclass[twocolumn,superscriptaddress,showpacs,nofootinbib,notitlepage,preprintnumbers,secnumarabic,amssymb, nobibnotes, aps, prd]{revtex4-2}
\usepackage[utf8]{inputenc}
\usepackage{graphicx}
\usepackage{latexsym,amsmath,amssymb,amsthm,lmodern,float,url,bm,IEEEtrantools}

\usepackage{listings}

\usepackage{mathtools}
\usepackage{natbib}
\usepackage{color}
\usepackage{microtype}
\usepackage{import}
\usepackage{bbold}
\usepackage[dvipsnames]{xcolor}
\usepackage[plain]{fancyref}
\usepackage{varioref}
\usepackage{slashed}
\usepackage{multirow}
\usepackage{tikz}
\usepackage{booktabs}
\usepackage{listings}
\usepackage{physics}
\usepackage{comment}
\usetikzlibrary{backgrounds}
\usetikzlibrary{arrows,shapes}
\usetikzlibrary{tikzmark}
\usetikzlibrary{calc}
\usetikzlibrary{positioning}
\usetikzlibrary{quantikz2}
\usepackage[normalem]{ulem}

\usetikzlibrary{positioning}
\usepackage{mathtools, nccmath}
\usepackage{wrapfig}
\usepackage{comment}
\usepackage{bbm}
\usepackage{adjustbox}

\usepackage[pdfusetitle]{hyperref}
\hypersetup{pdflang={English},colorlinks=true,linkcolor=RoyalBlue,citecolor=RoyalBlue,urlcolor=RoyalBlue} 
\usepackage[capitalise]{cleveref}

\usepackage{orcidlink}

\usepackage{todonotes}

\hypersetup{%
 bookmarksnumbered=true,
 pdftitle = {},
 pdfsubject = {},
 pdfauthor = {},
 pdfkeywords = {}
}

\newcommand{\fnal}{\affiliation{Fermi National Accelerator Laboratory, 
Batavia, IL 60510, USA}}

\begin{document}

\preprint{FERMILAB-PUB-26-0210-T}
\title{Ether of Orbifolds}

\author{Henry Lamm\,\orcidlink{0000-0003-3033-0791}}
\email{hlamm@fnal.gov}
\fnal

\date{\today}

\begin{abstract}
The orbifold lattice has been proposed as a route to practical quantum simulation of Yang--Mills theory, with claims of exponential speedup over all known approaches. Through analytical derivations, Monte Carlo simulation, and explicit circuit construction, we identify compounding costs entirely absent in Kogut--Susskind formulations: a mass-dependent Trotter overhead that scales as $m^4$, non-singlet contamination that grows as $m^2$ and worsens with penalty terms, and a mandatory mass extrapolation. Monte Carlo simulations of SU(3) establish a universal scaling: the continuum limit forces $m^2 \propto 1/a$, binding the Trotter step to the lattice spacing through a cost unique to orbifolds. For a fiducial $10^3$ calculation, the orbifold is $10^4$--$10^{10}$ times more expensive than every published alternative. These results indicate that the claimed computational advantages do not at present  survive quantitative scrutiny.
\end{abstract}

\maketitle

\section{Introduction}\label{sec:intro}

The quantum simulation community awaits a viable formulation of lattice QCD (LQCD) on quantum hardware~\cite{Bauer:2023qgm}. Over two decades since the first protocol~\cite{Byrnes:2005qx}, the field has not demonstrated quantum advantage for non-Abelian theories~\cite{Feynman:1981tf,Jordan:2012xnu,Zohar:2015hwa}. Many digitizations exist: electric basis~\cite{Byrnes:2005qx,Ciavarella:2021nmj}, discrete subgroups~\cite{Alexandru:2019nsa,Alexandru:2021jpm,Assi:2024pdn,Gustafson:2025s216}, loop-string-hadron~\cite{Raychowdhury:2019iki,Ciavarella:2025lsh}, $q$-deformed algebras~\cite{Zache:2023dko,Rhodes:2026atz}, mixed basis~\cite{Froland:2024suk}, qudit methods~\cite{Gustafson:2022xdt,Illa:2025qud}. There is no single best; each carries distinct trade-offs. The diversity is a strength. The LQCD program derives much of its credibility from the deliberate use of
diverse schemes. Multiple gauge actions each differing in their lattice
artifacts~\cite{Symanzik:1983dc,Iwasaki:1983iya,QCD-TARO:1999mox,Luscher:2009eq} are employed
alongside a variety of fermion formulations~\cite{Gattringer:2010zz}. Combined with different
lattice operators (e.g.\ conserved vs.\ local currents, smeared vs.\ thin links),
this pluralism ensures that when independent calculations agree on physical predictions, systematic
artifacts can be confidently excluded. The Flavour
Lattice Averaging Group exploit precisely this diversity, combining
results from collaborations worldwide to verify that the continuum,
infinite-volume, physical-mass limits are reliably
reached~\cite{FlavourLatticeAveragingGroupFLAG:2024oxs}.

A recent series of papers~\cite{Bergner:2024qjl,Halimeh:2024bth,Hanada:2025yzx,Bergner:2025zkj,Halimeh:2025ivn,Hanada:2024nqv} proposes the orbifold lattice as a ``universal framework'' with claims of exponential speedup over the Kogut--Susskind~(KS) formulation~\cite{Kogut:1974ag}. The central idea replaces the compact link variable $U \in \mathrm{SU}(N)$ with a noncompact complex $N\!\times\!N$ matrix $Z_{j,\vec{n}}$, decomposed as $Z = W\cdot U/\sqrt{2}$ where $W$ is positive-definite Hermitian and $U$ is unitary~\cite{Bergner:2025zkj,Buser:2020cvn,Kaplan:2002wv}. 
The Hamiltonian is $\hat{H}_\mathrm{orb} = \hat{K} + \hat{V}_\mathrm{plaq} + \hat{V}_D + \hat{V}_m + \hat{V}_\mathrm{U(1)}$, written in terms of rescaled dimensionless variables $Z_{j,\vec{n}}$ (absorbing $\sqrt{a^{d-2}/(2g^2)}$~\cite{Bergner:2025zkj,Buser:2020cvn}) and their $2N_c^2$ real conjugate momenta $\hat{p}_a$ per link in $d$ spatial dimensions. The first two terms mirror the KS Hamiltonian: $\hat{K}$ is the electric energy; $\hat{V}_\mathrm{plaq}$ is the magnetic energy:
\begin{align}
  \hat{K} &= \frac{g^2}{2a^{d-2}} \sum_{\ell,a} \hat{p}_a^2\\
  \hat{V}_\mathrm{plaq} &= \frac{1}{2g^2 a^{4-d}}\sum_{\square}\Tr\bigl|\hat{Z}_j\hat{Z}_k - \hat{Z}_k\hat{Z}_j\bigr|^2
  \end{align}
where the additional 3 terms enforce the reduction from the extended orbifold Hilbert space $\mathcal{H}_\mathrm{ext}$ to the target SU($N_c$) theory:\footnote{To be precise: all five terms are separately invariant under the orbifold gauge transformation $Z_{j,\vec{n}} \to \Omega_{\vec{n}}^{-1} Z_{j,\vec{n}} \Omega_{\vec{n}+\hat{j}}$~\cite{Kaplan:2002wv,Bergner:2024qjl}. The issue is not that the Hamiltonian breaks a gauge symmetry, but that $\mathcal{H}_\mathrm{ext}$ contains gauge-singlet states with no counterpart in SU($N_c$) Yang--Mills. These correspond to excitations of the radial mode $W$ and the U(1) phase. The terms below give these unphysical modes a large mass, so the low-energy spectrum approximates the target theory. We use ``non-singlet contamination'' and ``departure from SU($N_c$)'' to refer to the cost of this decoupling, not to a violation of the orbifold gauge Ward identity. One might, and we do often, colloquially refer to this issue though as gauge-violation.}
  \begin{align}
  \hat{V}_D &= \frac{g^2}{2a^{d}}\sum_{\vec{n}}\left|\sum_j\bigl(\hat{Z}_{j}\hat{\bar{Z}}_{j} - \hat{\bar{Z}}_{j{-}}\hat{Z}_{j{-}}\bigr)\right|^2\!,\label{eq:vd}\\
  \hat{V}_m &= \tfrac{1}{2}m^2 \sum_\ell \Tr\bigl(\hat{Z}_\ell \hat{\bar{Z}}_\ell - \tfrac{1}{2}\mathbf{1}\bigr)^2\,, \label{eq:vm}\\
  \hat{V}_\mathrm{U(1)} &= \tfrac{1}{2}m_\mathrm{U(1)}^2 \sum_\ell \bigl|2^{N_c/2}\!\det \hat{Z}_\ell - 1\bigr|^2\,.\label{eq:vu1}
\end{align}
$\hat{V}_D$ is the scalar kinetic D-term of the mother theory~\cite{Kaplan:2002wv}. The mass terms $\hat{V}_m$ and $\hat{V}_\mathrm{U(1)}$ carry no explicit $g$ or $a$ dependence in these variables, but $m$ itself must scale as $m^2 \gtrsim 1/a$ to suppress departures from the SU($N_c$) manifold. As $m,m_\mathrm{U(1)} \to \infty$, the radial mode $W \to \mathbf{1}$ and $\det U \to 1$, recovering the KS Hamiltonian; this limit already suggests the cost structure analyzed below.

Since all potentials are at most quartic in $Z$, the Hamiltonian decomposes into $\mathcal{O}(N_c^4 Q^4)$ Pauli strings per plaquette~\cite{Halimeh:2024bth} for $Q$-qubit encoding per boson, compared to $\mathcal{O}(2^Q)$ for na\"ive KS Pauli-string expansion. This is a per-step improvement the proponents characterize as exponential. Beyond gate counts, they argue that the orbifold resolves a deeper obstacle: constructing the gauge-invariant Hilbert space $\mathcal{H}_\mathrm{inv}$ for KS requires Clebsch--Gordan coefficients, multiplicity labels, and basis orthogonalization whose classical pre-processing cost grows exponentially with the number of qubits~\cite{Hanada:2025yzx}. The orbifold sidesteps this entirely by working in the extended Hilbert space $\mathcal{H}_\mathrm{ext}$, where every operator is a polynomial in $Z$ and $\bar{Z}$ with no group-theoretic compilation. Circuits for any SU($N_c$) in any dimension follow from this template. This is a genuine advantage: the orbifold provides analytically tractable Pauli-string Hamiltonians for arbitrary gauge group and dimension without group-theoretic compilation.

Such strong claims warrant careful quantitative examination. The approaches characterized as inferior in the orbifold literature deserve a thorough comparative analysis. Through analytic calculations, numerical simulations, and explicit circuits, we find that the path to practical quantum advantage via the orbifold formulation faces substantial obstacles not previously accounted for.

Further, the idea of replacing compact link variables with noncompact fields on the lattice has a long history. Early attempts using direct discretization of the continuum action without exact gauge invariance~\cite{Patrascioiu:1981ex,Stamatescu:1982qk,Seiler:1983mz,Seiler:1983rc,Cahill:1989jn} consistently found no confinement signal for SU(2) in 4d; Cahill's later work showed that imposing random compact gauge transformations restored the confinement signal for SU(2) and SU(3)~\cite{Cahill:1993ry,Cahill:1994gj,Cahill:1996ecg}, confirming that exact lattice gauge symmetry is essential. Palumbo~\cite{Palumbo:1990yp} showed that exact gauge invariance could be preserved with noncompact fields by embedding the link variables in $\mathrm{GL}(N,\mathbb{C})$ and introducing auxiliary fields, a construction whose algebraic structure closely parallels the orbifold decomposition $Z = W \cdot U$. Becchi and Palumbo~\cite{Becchi:1991tk,Becchi:1992jg} showed that the auxiliary-field coupling must be renormalized and runs to infinity in the continuum and computed $\Lambda_{\mathrm{NC}}/\Lambda_W$. This running is akin to requiring $m \to \infty$ as $a \to 0$. 

The Hamiltonian formulation~\cite{Diekmann:1993oe} recovers the Kogut-Susskind theory when the auxiliary coupling $\gamma \to \infty$, and one-loop effective Hamiltonian calculations~\cite{Borasoy:1994rh} quantified the ``$\gamma$-errors,'' i.e. lattice artifacts from the auxiliary field, with relative magnitude $\mathcal{O}(g_0^2/\tilde\gamma)$, where $\tilde\gamma = \gamma \cdot g_0$; these are direct analogues of the $\epsilon_g$ errors we study below. Monte Carlo studies found confinement consistent with Wilson's formulation~\cite{Palumbo:1992iw,DiCarlo:2000um} and showed that the noncompact regularization yields ${\sim}20\%$ larger physical volumes at matched lattice sizes, but requires simultaneous tuning of both the gauge coupling and the auxiliary mass. More recently, Babusci and Palumbo~\cite{Babusci:2024abc} derived modified Wilson actions by integrating out the auxiliary field, showing it produces a negative-definite correction. The orbifold formulation inherits this entire cost structure: the mass parameter $m$ plays the role of Palumbo's $\gamma$, and as we demonstrate quantitatively below, the requirement $m^2 \propto 1/a$ binds the Trotter overhead to the lattice spacing in a way that has no analogue in KS formulations. Bonati, Pelissetto, and Vicari~\cite{Bonati:2021jwt,Bonati:2021oss} have recently shown in a related context that gauge-symmetry breaking perturbations are RG-relevant at charged transitions, meaning they must be tuned to zero faster than a computable power of the correlation length which may suggest the orbifold's finite-$m$ artifacts are a relevant deformation rather than a benign correction.

\section{The mass--Trotter catastrophe}\label{sec:mass}

The orbifold lattice embeds SU($N$) into $\mathbb{R}^{2N^2}$ via complex link variables $Z = e^\phi U$. The mass term in Eq.~(\ref{eq:vm}) confines dynamics to the group manifold as $m \to \infty$. However, $m$ cannot be infinite in a quantum simulation, and the consequences of finite $m$ are more severe than previously recognized. We now examine each claimed advantage and quantify the associated costs.

\textit{The Trotter step scales with $m$.}---The natural Trotter splitting is $\hat{H} = \hat{K} + \hat{V}$ where $\hat{K} = \frac{1}{2}\sum_a\hat{p}_a^2$ and $\hat{V} = \hat{V}_\mathrm{plaq} + \hat{V}_D + \hat{V}_m + \hat{V}_\mathrm{U(1)}$. Since all potential terms are functions of the coordinates alone, they mutually commute; only the $\hat{K}$-$\hat{V}$ split produces Trotter error in this decomposition. For a second-order product formula, the number of Trotter steps to achieve total error $\leq \epsilon_T$ is~\cite{Childs:2021oam}
\begin{equation}
  r \sim \Bigl(\frac{t^3\,W_2}{\epsilon_T}\Bigr)^{\!1/2}\,,\quad W_2 = \|[V,[V,K]]\| + \tfrac{1}{2}\|[K,[K,V]]\|\,.
  \label{eq:pf2}
\end{equation}
Since $V = V_\mathrm{plaq} + V_D + V_m + V_\mathrm{U(1)}$ and the nested commutator $\|[V_i,[V_i,K]]\|$ scales as the square of the coupling in $V_i$, the mass term dominates at large $m$: $V_m$ contributes $\sim m^4$ per link while $V_\mathrm{plaq}$ contributes $\sim g^{-4} a^{2(d-4)}$, and recovering the target SU($N_c$) dynamics requires $m$ to be large. The mass-dominated potential gives $\|[V_m,[V_m,K]]\| \sim m^4 \cdot dV$ for $dV$ links, so the orbifold step count scales as
\begin{equation}
  r_\mathrm{orb} \sim \Bigl(\frac{t^3\,dV\,m^4}{\epsilon_T}\Bigr)^{\!1/2}\,.
  \label{eq:trotter_steps}
\end{equation}

The proponents~\cite{Hanada:2024nqv} claim a gentler scaling $r \propto m\,t$ based on numerical tests of the $\mathrm{S}^n$ model ($\hat{H} = \frac{1}{2}\sum\hat{p}_a^2 + \frac{m^2}{8}(\sum\hat{x}_a^2 - 1)^2$, a single link with no gauge coupling). However, their evidence has three significant limitations. (i)~The model contains \emph{no plaquette terms, no D-terms, and no inter-link coupling}. It is a single isolated link in the strong-coupling limit. The full Hamiltonian has non-commuting spatial interactions that contribute additional Trotter error independent of~$m$. (ii)~Every test evolves \emph{energy eigenstates}: ``We took the initial state to be the ground state of the truncated Hamiltonian''. For an eigenstate, the exact evolution is a trivial phase rotation $e^{-iE_k t}\ket{E_k}$ with no dynamical content e.g. no wavepacket spreading, no scattering, no thermalization. While Trotterization does introduce leakage to other eigenstates, this is a far gentler test than simulating genuine dynamics where accurate interference between many eigenstates over long times is required~\cite{Anand:2021xet,Gustafson:2026slb}. Passing the eigenstate test is necessary but insufficient to establish a Trotter step size for physically relevant simulations. (iii)~Near the ground state the quartic potential is effectively harmonic with $\omega \propto m$, and symplectic integrators for harmonic systems have the special property that accuracy depends on $\omega\!\cdot\!\Delta t$~\cite{RMcLachlan_1992,Hairer2006} --- a well-known artifact absent for generic anharmonic or interacting dynamics. While we will use this scaling as a lower bound when estimating gate costs, the applicability of this result to physically relevant simulations should be regarded with caution, and the more conservative bound of Eq.~(\ref{eq:trotter_steps}) should generally be preferred. 

\begin{figure*}[t]
\centering
\includegraphics[width=0.9\linewidth]{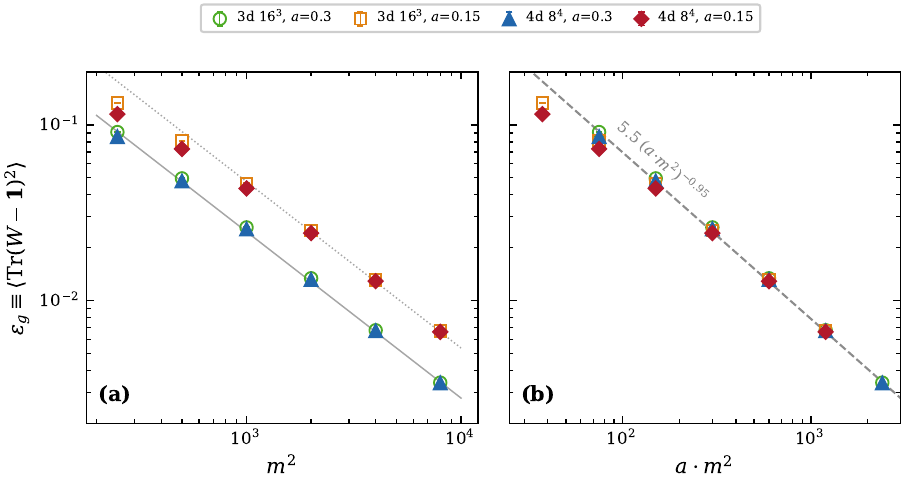}
\caption{(a)~Unitarity violation vs.\ $m^2$ for SU(3) orbifold MC on the largest volumes ($16^3$ for 3d, $8^4$ for 4d). Open markers: $(2{+}1)$d; filled: $(3{+}1)$d. Each color/symbol denotes a distinct (dimension, $a$) pair. Smaller volumes ($8^3$, $4^4$) agree to $<\!1\%$ (Table~\ref{tab:mc}). (b)~Scaling collapse: all four datasets plotted against $a\!\cdot\!m^2$. The dashed line shows $\propto\!(a\!\cdot\!m^2)^{-0.95}$.}
\label{fig:conv}
\end{figure*}

\textit{The mass needed for accuracy is large.}---At finite $m$, the link variables carry scalar excitations that must decouple. How large must $m$ be? We answer this with Monte Carlo simulations of the orbifold action for SU(3) in both $(2{+}1)$d and $(3{+}1)$d (the python code can be found in the supplemental information), at two lattice spacings in Table~\ref{tab:mc}.  The $(2{+}1)$d results were validated against the proponents' code~\cite{Bergner:2025zkj}. The key finding is a universal scaling (Fig.~\ref{fig:conv}b): all datasets collapse when plotted against $a\!\cdot\!m^2$, with $\epsilon_g\equiv\langle\Tr(W{-}\mathbf{1})^2\rangle \propto (a\!\cdot\!m^2)^{\alpha}$. The exponent $\alpha$ converges toward $-1$ at large $m^2$: fitting individual datasets at $m^2 \geq 2000$ gives $\alpha = -0.99\pm0.01$ ($a\!=\!0.3$) and $\alpha = -0.94\pm0.01$ ($a\!=\!0.15$) with $\chi^2/\mathrm{ndof} \sim 1$--$5$, while including $m^2 = 250$ pulls $\alpha$ to $-0.83$ due to sub-leading corrections from plaquette and D-terms. At the fiducial $m^2 \approx 8{,}400$ the scaling is effectively $\epsilon_g \propto 1/(a\!\cdot\!m^2)$. The controlling parameter is $a\!\cdot\!m^2$, not $m^2$ alone: maintaining a fixed departure from the SU($N_c$) manifold requires
\begin{equation}
    \label{eq:cae}
    m^2 \gtrsim C/(a\,\epsilon_g)\text{ with }C\approx 5\text{--}7,
\end{equation} so $a \to 0$ demands ever-larger $m$. Note that $\epsilon_g$ measures departure from the SU($N_c$) manifold, not a violation of the orbifold gauge symmetry, which is exact. The $(3{+}1)$d values are $1$--$4\%$ below $(2{+}1)$d at matched $a$, consistent with additional plaquette stiffness reducing radial fluctuations.

\begin{table*}[t]
\centering
\footnotesize
\setlength{\tabcolsep}{3pt}
\begin{tabular}{c|cccc|cccc|cc}
\hline\hline
 & \multicolumn{4}{c|}{$a = 0.3$} & \multicolumn{4}{c|}{$a = 0.15$} & \multicolumn{2}{c}{ratio ($\frac{a=0.15}{a=0.3}$)} \\
$m^2$ & 3d $8^3$ & 3d $16^3$ & 4d $4^4$ & 4d $8^4$ & 3d $8^3$ & 3d $16^3$ & 4d $4^4$ & 4d $8^4$ & 3d & 4d \\
\hline
250  & 0.0910(1) & 0.0910(1) & 0.0855(2) & 0.0858(1) & 0.1334(2) & 0.1332(1) & 0.1152(3) & 0.1151(1) & 1.47 & 1.34 \\
500  & 0.0497(1) & 0.0496(1) & 0.0477(1) & 0.0480(1) & 0.0811(1) & 0.0812(1) & 0.0733(2) & 0.0730(1) & 1.63 & 1.52 \\
1000 & 0.0261(1) & 0.0261(1) & 0.0254(1) & 0.0256(1) & 0.0465(1) & 0.0464(1) & 0.0432(1) & 0.0435(1) & 1.78 & 1.70 \\
2000 & 0.0134(1) & 0.0134(1) & 0.0132(1) & 0.0132(1) & 0.0251(1) & 0.0251(1) & 0.0242(1) & 0.0242(1) & 1.88 & 1.83 \\
4000 & 0.00678(1) & 0.00678(1) & 0.00675(2) & 0.00674(1) & 0.0131(1) & 0.0131(1) & 0.0128(1) & 0.0129(1) & 1.94 & 1.91 \\
8000 & 0.00341(1) & 0.00341(1) & 0.00340(1) & 0.00341(1) & 0.00672(1) & 0.00671(1) & 0.00666(2) & 0.00664(1) & 1.97 & 1.95 \\
\hline\hline
\end{tabular}
\caption{$\langle\Tr(W{-}\mathbf{1})^2\rangle$ for SU(3) orbifold lattice MC. Ratio columns show $a\!=\!0.15$ divided by $a\!=\!0.3$; both approach~2 at large $m^2$, confirming $a\!\cdot\!m^2$ scaling. All runs: $3{,}000$--$6{,}000$ Metropolis sweeps, $\ge\!500$ measurements per point.}
\label{tab:mc}
\end{table*}

\textit{KS has no analogous cost.}---To make the comparison explicit, consider the leading-order Trotter cost for both formulations in $d$ spatial dimensions with lattice volume $V = L^d$, $dV$ links, and $\sim\!d^2 V$ plaquettes. Let $Q$ denote the number of qubits per link. For KS with $K = \hat{H}_E$, $V = \hat{H}_B$ in Eq.~\eqref{eq:pf2}, the norms scale with the gauge coupling: at weak coupling, $\|\hat{H}_B\| \sim d^2 V/(g^2 a^{4-d})$ dominates and $W_2 \sim d^2 V\!\cdot\! g^{-4}\!a^{2(d-4)}$. The total KS cost is
\begin{equation}
  \mathcal{C}_\mathrm{KS} \sim f(Q)\,d^2 V \times \Bigl(\frac{t^3\,d^2 V}{g^4\,a^{2(4-d)}\,\epsilon_T}\Bigr)^{\!1/2}\,.
  \label{eq:cost_ks}
\end{equation}
This depends on $g$, $a$, and $d$, but on no mass or gauge-accuracy parameter. Using Eq.~\eqref{eq:trotter_steps} for the orbifold step count, the total orbifold cost is
\begin{equation}
  \mathcal{C}_\mathrm{orb} \sim N_c^4 Q^4\,d^2 V \times \Bigl(\frac{t^3\,dV\,m^4}{\epsilon_T}\Bigr)^{\!1/2} \times n_m\,,
  \label{eq:cost_orb}
\end{equation}
with $N_c$ the number of colors and $n_m \approx 5$ for multiple circuits required for mass extrapolation. Taking the ratio of Eq.~\eqref{eq:cost_orb} to Eq.~\eqref{eq:cost_ks} and substituting in Eq.~(\ref{eq:cae}):
\begin{equation}
  \frac{\mathcal{C}_\mathrm{orb}}{\mathcal{C}_\mathrm{KS}} \sim \frac{N_c^4\,Q^4}{f(Q)} \cdot \frac{C\,g^2}{\sqrt{d}\;\epsilon_g} \cdot n_m \cdot a^{3-d}\,.
  \label{eq:ratio}
\end{equation}
For $d = 3$, the lattice spacing $a$ drops out entirely: the ratio is set by $g^2/\epsilon_g$ --- the gauge coupling squared divided by the tolerated departure from unitarity. This is the central result. The orbifold's cost premium over KS is controlled by how close to the SU($N_c$) manifold ($\epsilon_g$) you demand, measured in units of $g^2$, times the per-step overhead $N_c^4 Q^4/f(Q)$ and the $n_m$ mass extrapolations. The per-plaquette cost $f(Q)$ for KS-type approaches is $\sim\!Q^2$ (quantum arithmetic~\cite{Kan:2021xfc}), $\sim\!Q$ (LCU~\cite{Rhodes:2024zbr}), or group-dependent (discrete subgroups). Note that $\epsilon_g$ is a systematic error, measuring the departure from the target SU($N_c$) theory, \emph{unique to orbifold}; its propagation to physical observable errors has not been quantified by the proponents.

\section{Non-singlet contamination \& penalty traps}\label{sec:gauge}

The orbifold Hamiltonian is gauge-invariant under the orbifold gauge group, but the extended Hilbert space $\mathcal{H}_\mathrm{ext}$ contains states with no counterpart in the target SU($N_c$) theory. On a finite grid, two distinct mechanisms populate these unphysical states, both of which worsen as $m$ increases. We quantify both for the $\mathrm{S}^1$ model (U(1) single link, two bosons $(x,y)$, gauge generator $\hat{G} = \hat{x}\hat{p}_y - \hat{y}\hat{p}_x$). The Hilbert space is discretized on a grid of $\Lambda$ points per boson in a box of half-width $R = 2$~\cite{Hanada:2024nqv}, giving $\Lambda^2$ states total. The full Hamiltonian and gauge generator were assembled as dense matrices and diagonalized exactly.

\textit{Truncation-induced symmetry breaking.}---In the continuum, $\hat{V}_m \propto (r^2 - 1)^2$ is radially symmetric and commutes exactly with $\hat{G}$. On a finite grid, the discrete Laplacian breaks rotational symmetry. We compute both the operator norm $\|[\hat{H}, \hat{G}^2]\|$ (worst-case non-singlet contamination over all states) and the ground-state expectation $\langle\hat{G}^2\rangle_\mathrm{g.s.}$ (non-singlet contamination of the lowest-energy state) as functions of both $m$ and $\Lambda$ in Table~\ref{tab:secIII}
\begin{table}
\caption{Non-singlet contamination for the $\mathrm{S}^1$ model (U(1) single link, $R\!=\!2$) as a function of $m$ and $\Lambda$. Left: operator norm $\|[\hat{H}, \hat{G}^2]\|$. Right: ground-state expectation $\langle\hat{G}^2\rangle_\mathrm{g.s.}$.}
\begin{center}
\setlength{\tabcolsep}{3pt}
\small
\begin{tabular}{r|ccc|ccc}
\hline\hline
 & \multicolumn{3}{c|}{$\|[\hat{H}, \hat{G}^2]\|$} & \multicolumn{3}{c}{$\langle\hat{G}^2\rangle_\mathrm{g.s.}$} \\
$m \backslash \Lambda$ & 20 & 32 & 40 & 20 & 32 & 40 \\
\hline
10 & $2\!\times\!10^4$ & $4\!\times\!10^4$ & $7\!\times\!10^4$ & $3\!\times\!10^{-3}$ & $4\!\times\!10^{-4}$ & $1.6\!\times\!10^{-4}$ \\
20 & $6\!\times\!10^4$ & $1\!\times\!10^5$ & $2\!\times\!10^5$ & $2\!\times\!10^{-2}$ & $4\!\times\!10^{-3}$ & $1.4\!\times\!10^{-3}$ \\
40 & $2\!\times\!10^5$ & $5\!\times\!10^5$ & $7\!\times\!10^5$ & $1\!\times\!10^{-1}$ & $3\!\times\!10^{-2}$ & $1\!\times\!10^{-2}$ \\
\hline\hline
\end{tabular}
\end{center}
\label{tab:secIII}
\end{table}
These two tables tell complementary stories. The right table shows that the ground-state non-singlet contamination $\langle\hat{G}^2\rangle_\mathrm{g.s.}$ \emph{decreases} with finer grids at fixed $m$, as expected: better resolution captures the singlet wavefunction more faithfully. However, at fixed $\Lambda$, increasing $m$ worsens $\langle\hat{G}^2\rangle_\mathrm{g.s.}$ because the narrowing wavefunction demands ever-finer resolution. This is a truncation cost: larger $m$ requires more qubits to maintain a given level of singlet accuracy.

Table~\ref{tab:secIII} reveals a more serious issue: the operator norm $\|[\hat{H}, \hat{G}^2]\|$ grows with \emph{both} $m$ and $\Lambda$. Finer grids expose higher-momentum modes with large non-singlet quantum numbers, and the $m^2$ prefactor in $\hat{V}_m$ amplifies their contribution. One might argue that low-energy states never populate these modes. But this assumes perfect state preparation and exact time evolution. In practice, Trotterization leaks $\mathcal{O}(\|[\hat{K},\hat{V}]\|\,\Delta t^3)$ amplitude per step into the full extended Hilbert space, including the UV sector where $\hat{G}^2$ is large. Each step pushes the state slightly away from singlets into an ever-growing reservoir of non-singlet modes. In KS, this problem does not arise: $[\hat{H}, \hat{G}] = 0$ exactly, so Trotter errors redistribute amplitude \emph{within} the gauge-invariant subspace but never leak out.

\textit{The penalty trap.}---Adding a penalty $\lambda\hat{G}^2$~\cite{Hanada:2024nqv} to suppress non-singlet contamination compounds the Trotter problem. We added $\lambda\hat{G}^2$ to $\hat{V}$ and recomputed $\|[\hat{K}, \hat{V}+\lambda\hat{G}^2]\|$ at $m=40$, $\Lambda = 20$ (Table~\ref{tab:penalty}).
\begin{table}[h]
\caption{Trotter commutator for the $\mathrm{S}^1$ model as a function of the penalty term coupling at $m\!=\!40$, $\Lambda\!=\!20$.}
\begin{center}
\begin{tabular}{c|cccc}
\hline\hline
$\lambda$ & 0 & 10 & 50 & 100 \\
\hline
$\|[\hat{K}, \hat{V}+\lambda\hat{G}^2]\|$ & $5.0\times 10^4$ & $7.1\times 10^4$ & $1.9\times 10^5$ & $3.3\times 10^5$ \\
\hline\hline
\end{tabular}
\end{center}
\label{tab:penalty}
\end{table}
Since $\hat{G}^2$ mixes position and momentum, it inflates $\|[\hat{K}, \hat{V}]\|$, requiring smaller $\Delta t$. The penalty intended to suppress non-singlet states increases Trotter errors, and therefore the very non-singlet leakage it was designed to suppress, creating a self-defeating feedback loop.

The orbifold thus faces a triple bind unique among all digitizations: increasing $m$ to decouple the unphysical modes (i)~requires finer grids to control ground-state non-singlet contamination, (ii)~enlarges the non-singlet UV reservoir that Trotter errors leak into, and (iii)~cannot be ameliorated by penalty terms without further inflating the Trotter commutator. No KS formulation shares any of these costs.

\section{Explicit circuits and T-gate costs}\label{sec:circuits}

We constructed explicit Trotter circuits for U(1) on a single link. Since the quartic potential produces $\mathcal{O}(Q^4)$ Pauli strings regardless of gauge group, the $m$-overhead applies universally, and for non-Abelian groups the $2N^2$ bosons per link would compound the orbifold cost.

\textit{Pauli-string decomposition of the orbifold potential.}---The orbifold uses $2Q$ qubits for two bosons $(x,y)$. In the coordinate basis with $2^Q$ grid points, $\hat{x} = \alpha_0 I + \sum_{j=1}^{Q}\alpha_j\hat{\sigma}_{z,j}$ is a sum of $Q$ single-$\sigma_z$ terms plus an identity shift. Since $\hat{\sigma}_{z}^2 = I$, the operator $\hat{x}^2$ reduces to Pauli strings of weight at most 2:
\begin{equation}
  \hat{x}^2 = c_0 I + \sum_j c_j \hat{\sigma}_{z,j} + \sum_{j<k} c_{jk}\hat{\sigma}_{z,j}\hat{\sigma}_{z,k}\,,
\end{equation}
giving $1 + Q + \binom{Q}{2}$ strings. For $\hat{x}^4 = (\hat{x}^2)^2$, products of pairs from $\hat{x}^2$ generate strings whose weight is the symmetric difference of the two factors' supports. Every subset of $\{1,\ldots,Q\}$ of size $\leq 4$ can be realized as such a symmetric difference, so
\begin{equation}
  \hat{x}^4 : \quad 1 + Q + \tbinom{Q}{2} + \tbinom{Q}{3} + \tbinom{Q}{4} \;\text{strings}\,.
  \label{eq:x4strings}
\end{equation}
For the cross-register term $\hat{x}^2\hat{y}^2$, each of the $(Q + \binom{Q}{2})$ nontrivial strings from $\hat{x}^2$ tensors with each of the $(Q + \binom{Q}{2})$ from $\hat{y}^2$, producing $(Q + \binom{Q}{2})^2$ truly cross-register strings (those acting nontrivially on both registers). These consist of $Q^2$ strings of weight 2 (one qubit per register), $2Q\binom{Q}{2}$ of weight 3, and $\binom{Q}{2}^2$ of weight 4.

The full potential $V = \frac{m^2}{8}(x^2 + y^2 - 1)^2$ has total nontrivial Pauli strings:
\begin{equation}
  N_\mathrm{pot} = 2\bigl[Q + \tbinom{Q}{2} + \tbinom{Q}{3} + \tbinom{Q}{4}\bigr] + \bigl(Q + \tbinom{Q}{2}\bigr)^{\!2}\,,
  \label{eq:npot}
\end{equation}
where the first bracket counts $x$-only and $y$-only strings from $x^4,y^4$ (which subsume those from $x^2,y^2$), and the second counts cross-register strings from $x^2y^2$.

\textit{$R_Z$ counting.}---Each weight-$w$ Pauli-$Z$ rotation $e^{-i\theta \hat{\sigma}_{z,i_1}\cdots\hat{\sigma}_{z,i_w}}$ requires $1$~$R_Z$ gate (plus Clifford conjugation). For a single application of $e^{-iV\Delta t}$, the $R_Z$ count equals the number of nontrivial Pauli strings $N_\mathrm{pot}$.

The kinetic term $K = \frac{1}{2}(p_x^2 + p_y^2)$ is diagonal in the momentum basis, contributing $2(Q + \binom{Q}{2})$ nontrivial strings. Switching bases requires a quantum Fourier transform (QFT) on each register: $\binom{Q}{2}$~$R_Z$ gates per QFT, with four QFTs total (forward and inverse on each register).

For the second-order Trotter $e^{-iV\Delta t/2}\,e^{-iK\Delta t}\,e^{-iV\Delta t/2}$, the total $R_Z$ count per step is:
\begin{equation}
  R_\mathrm{orb} = 2N_\mathrm{pot} + 2\bigl(Q + \tbinom{Q}{2}\bigr) + 4\tbinom{Q}{2}\,,\label{eq:rorb}
\end{equation}
where the three terms correspond to potential, kinetic diagonal, and QFT contributions.

In the electric basis $\ket{L}$, KS uses $Q$ qubits. The electric term $\frac{g^2}{2}\hat{L}^2$ is diagonal with $Q + \binom{Q}{2}$ nontrivial strings. The magnetic term $\cos\hat{\theta}$ requires a QFT, $Q$ $R_Z$ rotations, and inverse QFT. In the second-order splitting:
\begin{equation}
  R_\mathrm{KS} = 2\bigl(Q + \tbinom{Q}{2}\bigr) + 2\tbinom{Q}{2} + Q = 3Q + 4\tbinom{Q}{2}\,.\label{eq:rks}
\end{equation}
Each $R_Z$ is synthesized into $\sim\!1.15\log_2(1/\delta)$ T gates~\cite{CampbellETFQC}, where $\delta = \epsilon_T/(n_R \cdot r)$ distributes the total synthesis budget $\epsilon_T$ over all rotations. Table~\ref{tab:gates} compares the $R_Z$ counts per step and the resulting T-gate ratio.

\begin{table}[ht]
\centering
\caption{$R_Z$ gates per Trotter step, U(1) single link. Orbifold uses $2Q$ qubits; KS uses $Q$.}
\label{tab:gates}
\begin{tabular}{c|cccc}
\hline\hline
$Q$ & 4 & 5 & 6 & 8 \\
\hline
$R_{\mathrm{orb}}$ & 304 & 640 & 1{,}208 & 3{,}424 \\
$R_{\mathrm{KS}}$  & 36  & 55  & 78      & 136     \\
\hline\hline
\end{tabular}
\end{table}
The per-step $R_Z$ ratio grows as $\sim\!Q^2$, reflecting $\binom{Q}{2}^2 \sim Q^4$ cross-register strings versus $Q^2$ from KS. Since each $R_Z$ becomes $\sim\!30$--$80$ T gates after synthesis, and the orbifold requires far more Trotter steps, the T-gate gap compounds dramatically. Going from U(1) to SU($N_c$), the gap widens. For $SU(N_c)$, the orbifold uses $2N_c^2$ bosons per link. The quartic mass potential coupling all bosons generates $\sim N_c^4 Q^4$ Pauli strings per link. 

\section{Resource landscape}\label{sec:resources}

We assemble total T-gate estimates for a fiducial benchmark: pure-gauge Yang--Mills on a $10^3$ spatial lattice evolved for $10$~fm, with synthesis precision $\epsilon_T = 10^{-8}$, as proposed for shear viscosity~\cite{Cohen:2021imf}. For the orbifold we target $\epsilon_g = 0.01$ (1\% departure from unitarity), which from the scaling of Eq.~\ref{eq:cae} requires $m^2 \approx 8{,}400$. This forces $Q \geq 6$ ($\Lambda = 64$) to satisfy the proponents' own truncation criterion $\delta_x \leq 1/\sqrt{m}$~\cite{Hanada:2024nqv}. We take $n_m = 5$ and $R = 2$.

The key comparison is the Trotter step count. Established KS methods use $\Delta t \approx 0.2$~fm~\cite{Cohen:2021imf}. This is based on standard lattice field theory arguments about what should be necessary for suppressing temporal lattice artifacts~\cite{Carena:2021ltu}, giving $N_t = 50$ steps for $T = 10$~fm. The orbifold must evolve the same physical time but as discussed must work at smaller $\Delta t$. The extensive PF2 bound Eq.~\eqref{eq:trotter_steps} contains volume factors that make it a loose upper bound; extracting the per-link contribution gives $\Delta t \sim 1/m^2$, or $r \approx m^2 t \approx 8.4\times10^5$ steps while the proponents' empirical scaling $r \approx 6mt$ gives $\approx 5.5\times10^4$.

\begin{table}[t]
\caption{Total T-gate estimates for a fiducial calculation. Orbifold: $Q\!=\!6$, $a\!=\!0.1$~fm, $\epsilon_g\!=\!0.01$, $n_m\!=\!5$; the range spans the proponents' empirical $\Delta t \sim 1/m$ to the nested-commutator $\Delta t \sim 1/m^2$. Rows sorted by ascending cost. Several KS entries use truncations below what is needed for continuum-limit control; their costs will grow polynomially with finer truncation but remain well below the orbifold, which carries an additional $m$-dependent overhead absent in all KS schemes.}
\centering
\setlength{\tabcolsep}{2pt}
\footnotesize
\begin{tabular}{lcc|lcc}
\hline\hline
\multicolumn{3}{c|}{\emph{SU(2), $10^3$, $\epsilon_T\!=\!10^{-8}$}} & \multicolumn{3}{c}{\emph{SU(3), $10^3$, $\epsilon_T\!=\!10^{-8}$}} \\
Approach & T gates & Ref. & Approach & T gates & Ref.\\
\midrule
Triam.\ $j\!=\!\frac{1}{2}$$^\ddag$ & $\sim\!10^{7}$ & \cite{Kavaki:2024ijd}
  & Lg-$N_c$ LO$^\dag$ & $\sim\!10^{9}$ & \cite{Ciavarella:2025kry} \\
$\mathbb{BT}$ ord.\ prod. & $1.1\!\times\!10^{11}$ & \cite{Gustafson:2022xdt}
  & Elec. LCU$^\S$ & $10^{11}\text{--}10^{14}$ & \cite{Rhodes:2024zbr} \\
$\mathbb{BI}$ blk.\ enc. & $1.4\!\times\!10^{13}$ & \cite{Lamm:2024jnl}
  & Lg-$N_c$ NLO$^\dag$ & $\sim\!10^{12}$ & \cite{Ciavarella:2025kry} \\
Elec. Trotter & $3\!\times\!10^{19}$ & \cite{Kan:2021xfc}
  & $\Sigma(72\!\times\!3)$ & $3.5\!\times\!10^{12}$ & \cite{Gustafson:2025s216} \\
 & &
  & Elec. Trotter & $6.5\!\times\!10^{48}$ & \cite{Kan:2021xfc} \\
\hline
Orbifold (est.) & $10^{15}\text{--}10^{17}$ & $\star$
  & Orbifold (est.) & $10^{16}\text{--}10^{17}$ & $\star$ \\
\hline\hline
\multicolumn{6}{l}{\footnotesize$^\dag$Ref.~\cite{Ciavarella:2025kry} Table~V T/step $\times$ $N_t\!=\!50$.}\\
\multicolumn{6}{l}{$^\ddag$24 $R_Z$/cell $\times$ 125 cells $\times$ $N_t\!=\!50$. $\star$This work.}
\end{tabular}
\label{tab:resources}
\end{table}

Table~\ref{tab:resources} shows the orbifold is $10^{4}$--$10^{10}$ times more expensive than KS alternatives, with the range reflecting both uncertainty in the Trotter scaling and the diversity of KS approaches. For SU(2), the triamond lattice at $j_\mathrm{max} = 1/2$~\cite{Kavaki:2024ijd} achieves $\sim\!10^7$ T gates --- eight to ten orders of magnitude cheaper than the orbifold. For SU(3), the Krylov-subspace truncations of Ref.~\cite{Ciavarella:2025kry} require only $\sim\!10^9$--$10^{12}$ T gates depending on the truncation level. Even the most expensive KS entry ($\mathbb{BI}$ block encoding at $10^{13}$) is $10^2$--$10^4\times$ cheaper than the orbifold. The gap is driven by the three compounding factors unique to orbifold: the $m^4 \propto (a\,\epsilon_g)^{-2}$ contribution to the nested commutator Eq.~\eqref{eq:cost_orb}, the $N_c^4 Q^4$ per-step gate count, and $n_m \sim 5$ mass extrapolations. This overhead is not an artifact of product-formula simulation. Under qubitization or LCU methods, costs scale with the Hamiltonian 1-norm rather than nested commutators; the orbifold normalization factor $\lambda$ still inherits $m^2$-dependent mass terms absent in KS, reducing the scaling from $m^4$ to $m^2$ but preserving a qualitative gap tied to the mass parameter.

A fair criticism of Table~\ref{tab:resources} is that several KS entries operate at aggressive truncations ($j_\mathrm{max} = \tfrac{1}{2}$ for the triamond, $(1{,}1{,}1)$ Krylov subspace for the large-$N_c$ estimates) that are insufficient for controlled continuum-limit extrapolations. Enlarging these truncations will increase gate costs: the per-plaquette circuit depth grows polynomially with the truncation cutoff in every KS scheme listed. However, this growth does not close the gap with the orbifold. The discrete-subgroup approaches have been validated nonperturbatively through Euclidean Monte Carlo at multiple lattice spacings: S(1080) reproduces the SU(3) glueball spectrum to percent-level precision~\cite{Alexandru:2021jpm}, and $\Sigma(360\!\times\!3)$ exhibits Casimir scaling of the static potential~\cite{Assi:2024pdn}. The electric-basis LCU of Ref.~\cite{Rhodes:2024zbr} already spans truncations from minimal to large, with the reported $10^{11}$--$10^{14}$ range in Table~\ref{tab:resources} reflecting that variation; even the upper end remains orders of magnitude below the orbifold. Crucially, every KS entry in the table corresponds to an explicitly constructed circuit whose cost is known, whereas the orbifold estimate is derived from Pauli-string counting and has never been compiled to gates by any group. The point is not that current KS truncations are sufficient for physical predictions (they are not), but that their scaling with increasing truncation is polynomial and well characterized, whereas the orbifold carries an additional $m$-dependent cost floor that grows independently of and on top of any truncation refinement.

\section{On claims about the KS Hamiltonian}\label{sec:claims}

Several claims in the orbifold literature regarding the KS Hamiltonian warrant correction. Refs.~\cite{Bergner:2025zkj} state: ``unless using classical computers, one cannot even write the truncated Hamiltonian explicitly.'' This is contradicted by existing work: ~\cite{Balaji:2025afl,Balaji:2025yua} write down SU(3) circuits. Sparse-access oracles have been constructed for SU(2) and SU(3)~\cite{Rhodes:2024zbr}. Murairi et al.~\cite{Murairi:2022zdg} compiled complete Trotter circuits for two SU(2) truncations down to explicit CNOT counts. Rather than precomputing, Clebsch--Gordan coefficients are computable by quantum arithmetic~\cite{Kan:2021xfc,Ciavarella:2021nmj,Ciavarella:2025kry}. It is also noteworthy that despite the claim ``one can easily write the truncated Hamiltonian
explicitly by hand, and efficient quantum circuits can be designed by hand,'' no explicit orbifold circuit for any gauge group has appeared in~\cite{Bergner:2024qjl,Halimeh:2024bth,Hanada:2025yzx,Bergner:2025zkj,Halimeh:2025ivn,Hanada:2024nqv}. The circuits constructed in this work --- the first for the orbifold --- are not found to be efficient compared to KS alternatives.

The ``exponential speedup'' compares $\mathcal{O}(Q^4)$ to $\mathcal{O}(2^Q)$ for a na\"ive Pauli decomposition that no practitioner advocates. Multiple groups have independently constructed polynomial alternatives: Ciavarella et al.~\cite{Ciavarella:2021nmj} built exponentially better circuits for SU(3) plaquettes, Davoudi, Shaw, and Stryker~\cite{Davoudi:2022uzo} gave Trotter circuits scaling polynomially in $Q$, and Rhodes et al.~\cite{Rhodes:2024zbr} achieved $\mathcal{O}(Q)$ via LCU. More fundamentally, KS formulations yield local Hamiltonians, for which polynomial-overhead simulation has been known since Feynman~\cite{Feynman:1981tf} and Lloyd~\cite{Lloyd1073}. The claimed exponential advantage thus rests on a comparison with an approach that has been superseded. The dismissal of $\mathcal{H}_\mathrm{inv}$ as ``too costly''~\cite{Halimeh:2024bth,Hanada:2024nqv} ignores that KS eliminates the entire $m$-dependent cost structure dominating the orbifold budget. Truncations of SU(3) in KS formalism are remarkably efficient: Ciavarella, Burbano, and Bauer~\cite{Ciavarella:2025kry} achieve 1~qubit per plaquette with 17~$R_Z$ per plaquette in 2D and 157 in 3D at the $(1{,}1{,}1)$ truncation; Ciavarella and Bauer~\cite{Bauer:2024lkr} simulated SU(3) on $8\!\times\!8$ lattices on IBM hardware using this same truncation. To be clear, these implementations operate at truncations and volumes far below what is required for physical LQCD predictions. They are, however, demonstrative of the fact that KS-based circuits can be constructed and executed on present-day hardware, with clear roadmaps to larger truncations as quantum hardware scales.

\section{Discussion}\label{sec:conclusions}

The orbifold lattice provides analytically tractable Pauli-string Hamiltonians for any SU($N$), a genuine contribution. However, the costs of $m$-dependent Trotter overhead, non-singlet UV reservoir, the penalty trap, and mass extrapolation compound to orders-of-magnitude disadvantages. No head-to-head resource estimate against any modern KS implementation appears in Refs.~\cite{Halimeh:2024bth,Hanada:2025yzx,Bergner:2025zkj,Halimeh:2025ivn,Hanada:2024nqv,Bergner:2024qjl}.

Several questions must be addressed before the orbifold can be considered viable. (i)~A nonperturbative Euclidean calculation should reproduce known continuum results; notably, despite claims of universality, this work appears to be the first Monte Carlo study in 4d, and the same arguments for the orbifold's utility would seem to motivate Euclidean simulations as well. (ii)~The propagation of $\epsilon_g$ to physical observable errors must be quantified. Our Monte Carlo results provide a quantitative convergence rate to the full gauge group: the complex matrix $Z \in \mathbb{C}^{N\times N}$ approximates SU($N$) with error $\epsilon_g \sim C/(a\cdot m^2)$, but this convergence is entangled with both the continuum limit ($a\to 0$ demands larger $m$) and the truncation ($\Lambda$ must grow with $m$ to resolve the narrowing wavefunction), making the three limits mutually coupled rather than independently controllable. Investigations of even simple observables like string tension and glueball mass would provide valuable benchmarks. (iii)~Fermion-gauge coupling in the extended Hilbert space raises questions about renormalization, since the fermions couple to the full GL($N$,$\mathbb{C}$) link rather than an SU($N$) element. (iv)~A complete resource estimate must account for all systematics simultaneously.

With the history of LQCD as a guide, it is unlikely that one formulation will dominate. The classical program derives its credibility from diversity; the quantum program will be no different. Progress will require many approaches~\cite{Byrnes:2005qx,Ciavarella:2021nmj,Alexandru:2019nsa,Alexandru:2021jpm,Assi:2024pdn,Raychowdhury:2019iki,Zache:2023dko,Bergner:2024qjl,Kaplan:2002wv,Froland:2024suk,Rhodes:2026atz,Ciavarella:2025lsh}, of which orbifolds may prove to be one among several useful tools.

\begin{acknowledgments}
We thank M.~Hanada for communications including~\cite{Hanada:2026zab} received during revision of this paper. We thank M. Rhodes, A. Ciavarella, and E. Gustafson for constructive conversations and critical comments during the crafting of this paper. We thank the authors of Refs.~\cite{Halimeh:2024bth,Hanada:2024nqv,Bergner:2025zkj} for making their code and data publicly available. We disclose that this work was done with heavy collaboration with  \textsc{Claude} in producing the codes, writing, and reviewing. We acknowledge the support of the U.S. Department of Energy, Office of Science, Office of High Energy Physics Quantum Information Science Enabled Discovery (QuantISED) program ``Toward Lattice QCD on Quantum Computers" with E.G. under award number DE-SC0025940. This work was produced by Fermi Forward Discovery Group, LLC under Contract No. 89243024CSC000002 with the U.S. Department of Energy, Office of Science, Office of High Energy Physics.

\end{acknowledgments}

\bibliography{references}

\end{document}